\documentclass[10pt]{article}
\usepackage[OE]{express}
\usepackage{xcolor}
\usepackage{float}
\usepackage{multirow}

\begin{document}
\title{Fully integrated free-running InGaAs/InP single-photon detector for accurate lidar applications}

\author{Chao Yu,\authormark{1,2,4} Mingjia Shangguan,\authormark{1,3,4} Haiyun Xia,\authormark{2,3,5} Jun Zhang,\authormark{1,2,6} Xiankang Dou,\authormark{3} and
Jian-Wei Pan,\authormark{1,2}}

\address{\authormark{1}Hefei National Laboratory for Physical Sciences at the Microscale and Department of Modern Physics, University of Science and Technology of China, Hefei, Anhui 230026, China\\
\authormark{2}CAS Center for Excellence and Synergetic Innovation Center in Quantum Information and Quantum Physics, University of Science and Technology of China, Hefei, Anhui 230026, China\\
\authormark{3}CAS Key Laboratory of Geospace Environment, University of Science and Technology of China, Hefei, Anhui 230026, China\\
\authormark{4}These authors contributed equally to this work\\
\textcolor{blue}{\authormark{5}hsia@ustc.edu.cn}\\
\textcolor{blue}{\authormark{6}zhangjun@ustc.edu.cn}}






\begin{abstract}
We present a fully integrated InGaAs/InP negative feedback avalanche diode (NFAD) based free-running single-photon detector (SPD) designed for accurate lidar applications.
A free-piston Stirling cooler is used to cool down the NFAD with a large temperature range, and an active hold-off circuit implemented in a field programmable gate array
is applied to further suppress the afterpulsing contribution. The key parameters of the free-running SPD including photon detection efficiency (PDE), dark count rate (DCR), afterpulse probability, and maximum count rate (MCR) are dedicatedly optimized for lidar application in practice. We then perform a field experiment using a Mie lidar system with 20 kHz pulse repetition frequency to compare the performance between the free-running InGaAs/InP SPD and a commercial superconducting nanowire single-photon detector (SNSPD). Our detector exhibits good performance with 1.6 Mcps MCR (0.6 $\mu$s hold-off time), 10\% PDE, 950 cps DCR, and 18\% afterpulse probability over 50 $\mu$s period. Such performance is worse than the SNSPD with 60\% PDE and 300 cps DCR. However, after performing a specific algorithm that we have developed for afterpulse and count rate corrections, the lidar system performance in terms of range-corrected signal ($Pr^2$) distribution using our SPD agrees very well with
the result using the SNSPD, with only a relative error of $\sim$2\%. Due to the advantages of low-cost and small size of InGaAs/InP NFADs, such detector provides a practical solution for accurate lidar applications.
\end{abstract}



\section{Introduction}

Aerosol lidars play an important role in many fields, such as atmosphere environment management, agricultural meteorology, hydrological cycle and radiation budget. Compared with ultraviolet (UV) and Visible systems, 1.5 $\mu$m aerosol lidars offer several advantages, including higher maximum permissible exposure to human eyes, lower atmospheric attenuation, minor disturbance from Rayleigh scattering, and weaker sky radiance. For instance, when using a multifrequency lidar for retrieving the spatial distribution of respirable fractions of aerosol in the lower atmosphere, 1.5 $\mu$m aerosol lidars provide more reliable concentration estimates of coarse aerosol particles~\cite{Lisenko16}.

Using single-photon detectors (SPDs) at 1.5 $\mu$m can greatly improve the performance of aerosol lidars. Currently the main techniques for single-photon detection at the telecom wavelength~\cite{Eisaman11} include superconducting nanowire single-photon detectors (SNSPDs), up-conversion SPDs and InGaAs/InP single-photon avalanche diodes (SPADs). SNSPDs exhibit excellent performance of high photon detection efficiency (PDE), low dark count rate (DCR) and low timing jitter~\cite{MVS13,YLZ14,WYC17}, however, the requirement of cryogenic conditions limits the use for practical applications. Up-conversion SPDs exhibit moderate performance, and recently have been used for lidar experiments~\cite{Xia15,Hogstedt16,Shangguan16,XSW16}. Compared with the above detectors, InGaAs/InP SPADs have relatively poor performance, however, such detectors are widely used in practical applications due to their advantages of small size and low cost~\cite{Itzler11,Zhang15}. The key parameters of InGaAs/InP SPADs, i.e., PDE, DCR, afterpulse probability and maximum count rate (MCR), influence each other, so that these parameters should be compromised and optimized according to the requirements of applications. Apart from SPAD device, the quenching electronics is the crucial part for a SPAD~\cite{Zhang15}. InGaAs/InP SPADs are often operated either in gating mode or in free-running mode, which are suited for synchronous and asynchronous single-photon detections, respectively.

Gating mode can effectively reduce DCR of InGaAs/InP SPADs, and gating frequencies have been increased from a few MHz in the early stage to the regime of GHz~\cite{Zhang15}. High-frequency gating techniques, including sine wave gating~\cite{NSI06,GAP09,NAI09,GAP10,NTY11,LLW12,GAP12,RBM13} and self-differencing~\cite{Toshiba07,CWW10}, can highly shorten avalanche duration time and thus significantly suppress the afterpulsing effect, which greatly helps to increase count rates of InGaAs/InP SPADs.
However, gating mode is not well suited for remote sensing applications due to the limit of small duty cycle for asynchronous single-photon detection~\cite{ELZ10,RGL11}.
Even using high-frequency gating technique, the duty cycle can reach around 20\%, which still results in pretty low effective detection efficiency~\cite{Zhang15}.

For lidar and remote sensing applications~\cite{Xia07,Xia12,Xia14,Xia16}, the most practical solution is using free-running InGaAs/InP SPADs. So far several techniques have been used to implement free-running operations. First, passive quenching is certainly a common approach for free-running mode.
Rarity et al. demonstrated a passively quenched InGaAs/InP SPD in 2000~\cite{RWR00}. The important technical issue to be settled in such detectors is to suppress afterpulsing effect~\cite{Zhang15}. Using long recovery time can reduce the afterpulse probability. However, the performance of count rate is severely limited. Warburton et al. presented a solution for free-running operation through lowering excess bias and increasing temperature~\cite{WIB09}. The Virginia group demonstrated a scheme called passive quenching and active reset (PQAR)~\cite{LHC08,HLZ10}, combining the advantages of
parasitic capacitance minimization by chip-to-chip wire bonding and fast electronic switch to swiftly reset the bias voltage after the hold-off time.
Second, active quenching can also be used for free-running mode given a short quenching time.
The Geneva group demonstrated a practical free-running detector using integrated circuit of active quenching, in which both the parasitic capacitance and the propagation delay of the quenching circuit were minimized~\cite{TSG07,ZTG09}. Third, new SPAD devices called negative feedback avalanche diodes (NFADs) provide a simple and practical solution for free-running mode~\cite{IJN09,IJO10,JIO11}. NFADs monolithically integrate thin film resisters ($\sim$ 900 K$\Omega$) inside the semiconductor structure of devices for passive quenching operation, which can fundamentally solve the problem of parasitic capacitance compared with the PQAR scheme. Yan et al. implemented a NFAD based free-running detector
with 10\% PDE and 100 cps DCR at 193 K~\cite{YHH12}. Korzh et al. demonstrated a similar detector and optimized the operation conditions to further reduce DCR down to 1 cps at 10\% PDE~\cite{KWL14}. However, these two free-running SPDs used pretty long hold-off time to further suppress the afterpulse probability, which results in
low MCR and even signal distortion in lidar applications.

In this paper, we present a fully integrated free-running SPD based on InGaAs/InP NFAD (Princeton Lightwave) with 1.6 Mcps MCR, 10\% PDE and 950 cps DCR. We then develop a specific algorithm for afterpulse correction and count rate correction for lidar applications. After the corrections, the lidar system performance in terms of range-corrected signal ($Pr^2$) distribution using such detector is considerably comparable to that of using a commercial SNSPD with only a relative error of $\sim$2\%.

\section{Mie lidar at 1.5 $\mu$m}

\begin{figure}[tbp]
\centering
\includegraphics[width=8.8 cm]{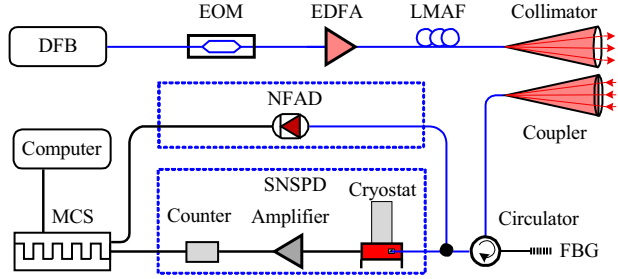}
\caption{Experimental setup of aerosol lidar using free-running InGaAs/InP single-photon detector. DFB: distributed feedback diode; EOM: electro-optic modulator; EDFA: erbium-doped fiber amplifier; LMAF: large-mode-area fiber; FBG: fiber bragg grating; NFAD: negative feedback avalanche diode; SNSPD: superconducting nanowire single-photon detector; MCS: multi-channel scaler.}
\label{fig1}
\end{figure}

The experimental setup of 1.5 $\mu$m Mie lidar using free-running InGaAs/InP SPD is shown in Fig.~\ref{fig1}. The whole laser system of lidar utilizes master-oscillator power-amplifier architecture. A continuous wave (CW) laser from a distributed feedback diode (DFB, 1548.1 nm) is chopped into a pulse train using an electro-optic modulator (EOM, Photline, MXER-LN-10) with high extinction ratio (35 dB). The EOM is driven by a pulse generator, which controls the shape and pulse repetition frequency (PRF) of the laser. In the experiment, the PRF of the pulses is set to 20 kHz, which indicates the maximum unambiguous detection range is $\sim$7.5 km.
 The weak laser pulses are fed into an erbium-doped fiber amplifier (EDFA, Keyopsys, PEFA-EOLA), which emits a pulse train with 110 $\mu$J pulse energy and 200 ns pulse width. A large-mode-area fiber with numerical aperture of 0.08 is used to increase the threshold of stimulated Brillouin scattering and to avoid self-saturation of amplified spontaneous emission (ASE). The laser is then collimated and sent to the atmosphere.

The backscattering signal from the atmosphere is collected into a single-mode fiber using a pigtailed coupler. The background noise is filtered out by inserting a fiber Bragg grating (FBG, Advanced Optics Solution GmbH) with a ultra-narrow bandwidth of 6 pm and an insertion loss less than 3 dB. The backscattering signal is detected by an InGaAs/InP NFAD based free-running SPD. The detection signals of the received photons are recorded on a multi-channel scaler (FAST ComTec, MCS6A) and then transmitted to a computer.
In order to optimize the performance of free-running InGaAs/InP SPD for lidar applications, a commercial superconductor nanowire single photon detector (SNSPD, Single Quantum, Eos 210 CS) is used to detect the atmospheric backscattering signal simultaneously for comparison.
The SNSPD exhibits pretty good calibration performance with 60\% PDE, 300 cps DCR, low timing jitter and high MCR. Therefore, such detector is well suited to be used as a reference in the experiment.

\section{Free-running InGaAs/InP single-photon detector}

\begin{figure}[tbp]
\centering
\includegraphics[width=11.6 cm]{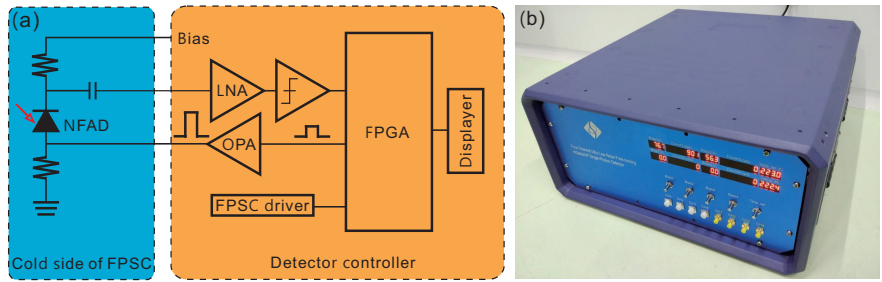}
\caption{Design diagram (a) and photo (b) of the free-running InGaAs/InP single-photon detector system. LNA: low-noise amplifier; OPA: operational amplifier; FPGA: field-programmable gate array; FPSC: free-piston Stirling cooler.}
\label{fig2}
\end{figure}

Backscattering signal from a lidar decays very fast along detection range so that the count rate of signal has a large dynamic range. Therefore, a SPD with high MCR is required to avoid detector saturation, particularly in the near range. The main design considerations of our free-running InGaAs/InP detector system include moderate DCR and afterpulse performance with high MCR and integrated module for practical use.
The design diagram and photo of the detector are shown in Figs.~\ref{fig2}(a) and \ref{fig2}(b), respectively. To achieve low DCR performance, cooling down the NFAD device is crucial.
In the detector, a free-piston Stirling cooler (FPSC, SC-UE15R) is used, which has a powerful cooling capacity and can easily cool down the NFAD to 163 K. The NFAD is fixed on the cold side surface of FPSC and then the whole cold side is encapsulated inside a cavity. The cavity is connected with the detector controller via a micro rectangular connector for temperature control and micro coaxial (MCX) connectors for electronic signal transmission. A target temperature is set by rotating a potentiometer and displayed on the detector panel. Since the cooling power of FPSC is tuned by a direct current (DC) voltage in its driver circuit, a proportional-integral-derivative (PID) program is developed in a field-programmable gate array (FPGA) to regulate the voltage in real-time. In such a way, the cooling temperature that is displayed on the detector panel in real-time can be finally stabilized at the target temperature with an error less than 1 K.

The reverse bias voltage of NFAD is tuned by another potentiometer on the detector panel to guarantee that the NFAD is operated in Geiger mode. Once an avalanche occurs, the original avalanche signal is alternating current (AC) coupled to a low-noise amplifier (LNA) with a gain of 40 dB. The amplified signal is then discriminated. The output signals of the discriminator are sent to the FPGA for further processing. The FPGA counts the discriminated signals and drivers the displayer on the panel, and outputs reset pulses as well. The reset LVTTL pulses are amplified to 5 V via an operational amplifier (OPA) and then coupled to the anode of NFAD. When the reset pulses are at HIGH level, the bias is below breakdown voltage and hence the width of the reset pulses are exactly the hold-off time. Once the status of the reset pulses is switched from HIGH level to LOW level, the NFAD is directly reset to the initial state. Due to the capacitance response of NFAD, the falling edges of reset pulses produce negative derivative signals that are superposed with avalanche signals. In order to avoid wrong detections, the output of the discriminator is latched until 20 ns after the falling edges of reset pulses so that negative derivative signals are ignored.

\section{Optimization of parameters}

\begin{figure}[tbp]
\centering
\includegraphics[width=10.8 cm]{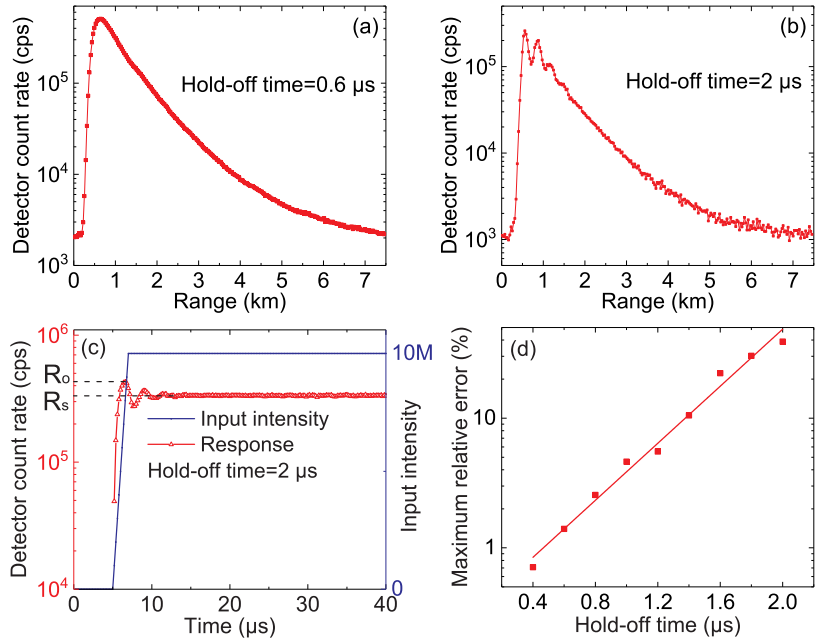}
\caption{Experimental backscattering signal with hold-off time of 0.6 $\mu$s (a) and 2 $\mu$s (b) in the InGaAs/InP single-photon detector. (c) Simulation illustration of the ringing effect. It is assumed that the input intensity is a step signal from 0 to $10^7$ photons with a rising time of 2 $\mu$s and the hold-off time of detector is 2 $\mu$s. (d) The plot of maximum relative overshoot error as a function of hold-off time.}
\label{fig3}
\end{figure}

Afterpulse probability is an additional parameter to be considered in lidar applications compared with SNSPD. This parameter is related to the operation
conditions of NFAD including excess bias voltage, temperature, and hold-off time. Setting a long hold-off time can effectively reduce aftepulse probability,
which, however, considerably limits the parameter of MCR and results in a ringing effect to the detected backscattering signal in lidar applications.
Figures~\ref{fig3}(a) and \ref{fig3}(b) show the backscattering signals detected by the free-running InGaAs/InP SPD with different settings of hold-off time. In the case of 2 $\mu$s hold-off time one can clearly see the ringing effect of detector count rate in the peak area while the ringing disappears in the case of 0.6 $\mu$s hold-off time.

In order to illustrate the ringing effect, we develop a MATLAB program to simulate the detector count rate of free-running InGaAs/InP SPD with different settings of hold-off time. The input intensity is assumed to be a step signal from 0 to $10^7$ photons with a rising time of 2 $\mu$s, which are close to the values in the field experiment. Given a setting of 2 $\mu$s hold-off time, the ringing effect distinctly appears, as shown in Fig.~\ref{fig3}(c), from which
one can conclude that such effect highly relies on the input photon flux and the hold-off time setting of SPD.
Further, let us define the maximum count rate as $R_o$ and the stable count rate as $R_s$, the relationship between $R_s$ and the input intensity is given by
\begin{equation}
\label{Rstable}
I\eta=\frac{R_{s}}{1-R_{s}\tau}-DCR.
\end{equation}
where $\eta$ is the detection efficiency, $\tau$ is the hold-off time, and $I$ is the input photon flux. The maximum relative overshoot error is calculated as
\begin{equation}
\label{Err}
Err_{os}=\frac{R_{o}-R_{s}}{R_{s}}.
\end{equation}
Figure~\ref{fig3}(d) plots $Err_{os}$ as a function of hold-off time, which indicates that such overshoot error exponentially increases with the increase of $\tau$. Therefore, for lidar applications using InGaAs/InP SPD the hold-off time has to be chosen as small as possible to avoid the count rate ringing. In our experiment, the hold-off time is set as 0.6 $\mu$s, corresponding to a MCR of $\sim$1.6 Mcps, to guarantee that the maximum relative overshoot error is less than 1.5\%.

\begin{figure}[tbp]
\centering
\includegraphics[width=8.8 cm]{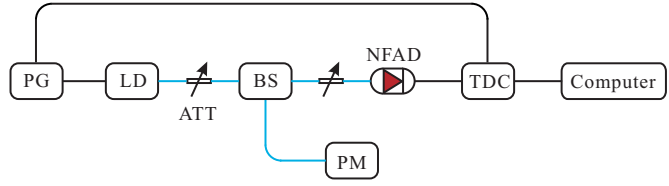}
\caption{Experimental setup for characterizing the free-running InGaAs/InP SPD. PG: pulse generator; LD: laser diode; PM: power meter; ATT: attenuator; TDC: time-to-digital converter.}
\label{fig4}
\end{figure}

We then calibrate the key parameters of the free-running InGaAs/InP SPD with 0.6 $\mu$s hold-off time including PDE, DCR, and afterpulse probability distribution.
The setup for calibration is shown in Fig.~\ref{fig4}. A laser diode (LD, picoquant PDL 800-D) is driven by a pulse generator (PG) with a frequency of 20 kHz that is the same as the PRF of our lidar system. The laser pulses are split by a beam splitter (BS). One output port of the BS is monitored by a power meter (PM, Exfo IQS-1600), and the other one is connected with a precise attenuator (ATT) to further attenuate the intensity down to single-photon level. The detector outputs are recorded by a time-to-digital converter (TDC, picoquant PicoHarp300), which is triggered by synchronized signals from the PG.

For DCR measurement, the LD is switched off. Given a count rate $R$, the inherent dark count rate $r$ is calculated as
\begin{equation}
\label{r}
r_d=\frac{R}{1-R\tau}.
\end{equation}
When the LD is switched on, the peak photon count rate $S_1$ and total count rate $S_2$ are recorded, respectively. Then PDE is calculated as
\begin{equation}
\label{pde}
PDE=-\frac{1}{\mu}\ln(1-\frac{S_1}{f}),
\end{equation}
where $\mu$ is the mean photon number per pulse, $f$ is the frequency of laser pulses. The distribution of afterpulse counts can be obtained from the TDC, and the total afterpulse probability over the whole 50 $\mu$s period is evaluated by
\begin{equation}
\label{pap}
P_{ap}=\frac{S_2-S_1-r_d(1-S_2\tau)}{S_1}.
\end{equation}

\begin{table}[tbp]
\centering
\caption{The calibration results of dark count rate and afterpulse probability at different temperatures (T) with 10\% PDE.}
\begin{tabular}{|l|l|l|l|}
\hline
T (K)   &DCR (cps)   &$P_{ap}$ (\%) \\ \hline
243   &6743 &13.2\\ \hline
223  &952 &18.0\\ \hline
203  &135 &38.5\\ \hline
\end{tabular}
\label{table1}
\end{table}

\begin{figure}[tbp]
\centering
\includegraphics[width=6.5 cm]{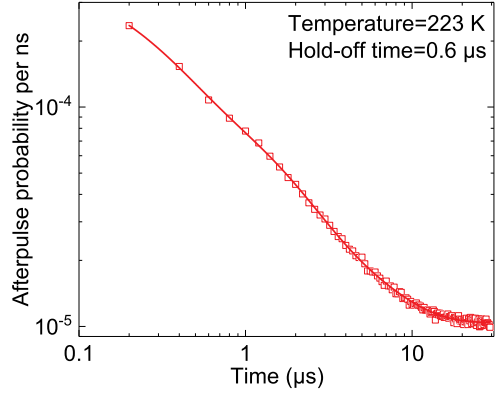}
\caption{Experimental result (square symbol) and fitted curve (line) of afterpulse probability distribution with 10\% PDE and 0.6 $\mu$s hold-off time at 223 K.}
\label{fig5}
\end{figure}

Table~\ref{table1} shows the calibration results of DCR and afterpulse probability with 10\% PDE at three typical temperatures with
hold-off time of 0.6 $\mu$s. At low temperatures, the SPD exhibits low DCR performance but high afterpulse probability. Therefore, these two parameters have to be compromised for lidar applications. In the lidar experiment, the temperature of the SPD is set as 223 K with performance of 10\% PDE, 950 cps DCR and 18\% afterpulse probability.
Such performance is moderately better than the commercial product of free-running InGaAs/InP SPD (ID220-FR-SMF, IDQ), which has 1 kcps DCR and 10\% PDE with
hold-off time of 10 $\mu$s~\cite{IDQ}.
The afterpulse probability distribution of our free-running SPD is further measured using the TDC, as shown in Fig.~\ref{fig5}. This distribution is fitted by the following formula~\cite{ZTG09}
\begin{equation}
\label{pap_t}
P_{ap}(t)=A_1Exp(-t/\tau_1)+A_2Exp(-t/\tau_2)+A_3Exp(-t/\tau_3)+c,
\end{equation}
where $A_1, \tau_1, A_2, \tau_2, A_3, \tau_3, c$ are fitting parameters. Acquiring such distribution is useful for afterpulse correction in the lidar experiment.

\section{Afterpulse and count rate corrections for lidar experiment}

\begin{figure}[tbp]
\centering
\includegraphics[width=10.5 cm]{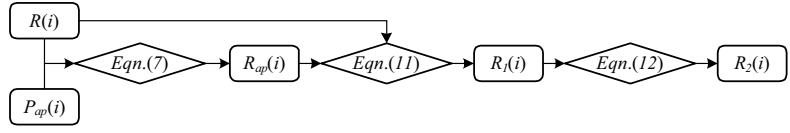}
\caption{Flow diagram of the correction algorithm. $R(i)$, $P_{ap}(i)$ represent the measured detector count rate, afterpulse probability in bin $i$, respectively.
$R_{ap}(i)$, $R_1(i)$ and $R_2(i)$ represent the calculated afterpulse count rate, count rate with afterpulse correction, and count rate with both corrections in in bin $i$, respectively.}
\label{fig6}
\end{figure}

In order to apply our free-running InGaAs/InP SPD to practical lidar system, we develop a specific algorithm for afterpulse and count rate corrections, after which the lidar performance has been significantly improved. The flow diagram of the correction algorithm is described in Fig.~\ref{fig6}. Given a measured afterpulse probability distribution $P_{ap}(i)$ and backscattering signal count rate distribution $R(i)$, the afterpulse count rate $R_{ap}$ in bin $j$ is
\begin{equation}
\label{Rap}
R_{ap}(j)=\sum_iA(i,j),
\end{equation}
where $A(i,j)$ represents the afterpulse count rate probability in bin $j$ due to a detection click in bin $i$. $A(i,j)$ can be calculated as
\begin{equation}
\label{Aij}
A(i,j)=R(i)P_{nc}(i,j)P_{nap}(i,j)P_{ap}(j;i),
\end{equation}
where $P_{nc}(i,j)$ represents the probability that no photon count occurs between bin $i$ and bin $j$, $P_{nap}(i,j)$ represents the probability that no afterpulse occurs between bin $i$ and bin $j$, and $P_{ap}(j;i)$ represents the probability that a photon count in bin $i$ results in an afterpulse count in bin $j$. $P_{nc}(i,j)$ and $P_{nap}(i,j)$ are further calculated by
\begin{equation}
\label{Pnc}
P_{nc}(i,j)=Exp[-\sum_{k=i}^jR(k)bin_w],
\end{equation}
\begin{equation}
\label{Pnap}
P_{nap}(i,j)=Exp[-\sum_{k=0}^{j-i-1}P_{ap}(k)],
\end{equation}
where $bin_w$ is the duration time of a bin. Therefore, the count rate in bin $i$ without the afterpulsing contribution is given by
\begin{equation}
\label{R1}
R_1(i)=R(i)-R_{ap}(i),
\end{equation}
and further considering the limit of hold-off time the corrected photon count rate in bin $i$ is
\begin{equation}
\label{R2}
R_2(i)=\frac{R_1(i)}{1-R(i)\tau}-DCR(i).
\end{equation}

\begin{figure}[tbp]
\centering
\includegraphics[width=10.5 cm]{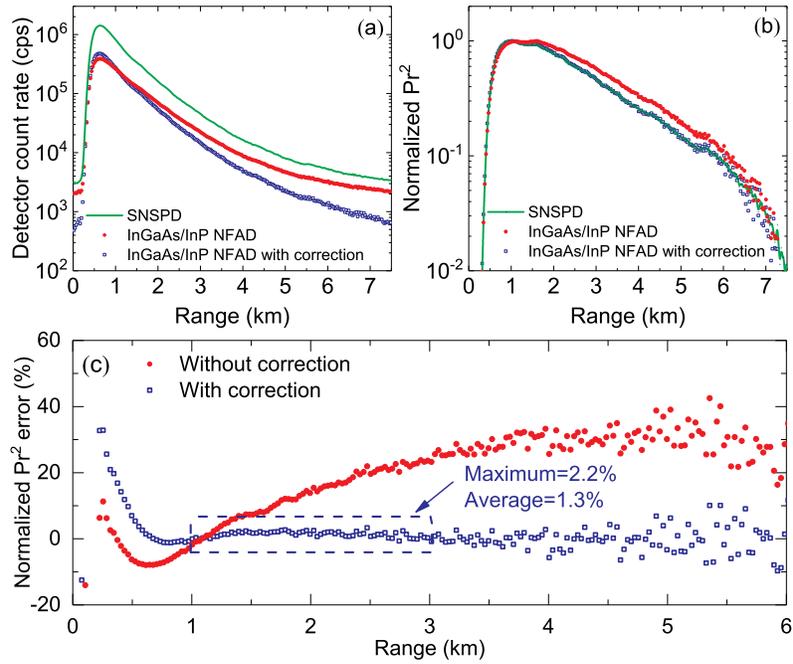}
\caption{The detector count rate (a) and normalized $Pr^2$ (b) as a function of range in the field lidar experiment using SNSPD (green lines), and InGaAs/InP NFAD (red circle symbols), respectively. The results using InGaAs/InP NFAD are further corrected (blue square symbols) by the afterpulse and count rate correction algorithm. (c) Given the measured results using SNSPD as a reference, the relative error of normalized $Pr^2$ as a function of range using InGaAs/InP NFAD without (red circle symbols) and with (blue square symbols) correction.}
\label{fig7}
\end{figure}

In the field lidar experiment, the measured results of detector count rate and range-corrected signal (normalized $Pr^2$) versus range using both SNSPD and InGaAs/InP NFAD are shown in Figs.~\ref{fig7}(a) and \ref{fig7}(b), respectively. After correcting the results using InGaAs/InP NFAD by the afterpulse and count rate correction algorithm,
one can find out that the lidar performances in terms of normalized $Pr^2$ versus range in two cases of using SNSPD and InGaAs/InP NFAD with the corrections agree very well with each other, which clearly shows the usefulness of the correction algorithm. Further, we use the normalized $Pr^2$ results using SNSPD as a reference and calculate the relative error of the results using InGaAs/InP NFAD, as shown in Fig.~\ref{fig7}(c). In the case without correction, the relative error between the two scenarios increases and then is stable at around 30\% with the increase of range. However, with correction such relative error is drastically decreased down to a pretty low level. For instance, in the range between 1 km to 3 km, the maximum error is only 2.2\% while the average error is 1.3\%. The large errors in the range close to the limit, i.e., $\sim$6 km, are due to the fact that detector count rate significantly decreases in that range and thus the calculation of relative errors is not accurate because of statistical fluctuations.
Due to the geometrical overlap factor in the biaxial lidar, the raw signal increases rapidly from zero to maximum in the near range. Since the scaler receives raw signals from two detectors, a slight delay between two channels results in very large errors due to the steep slopes in the very near range, i.e., < 0.5 km. In practice, the results of normalized $Pr^2$ in very near range are not used.
Such comparison indicates that for lidar applications our free-running InGaAs/InP SPD is completely comparable to SNSPD with the help of afterpulse and count rate corrections.

\section{Conclusion}
In conclusion, we have presented a fully integrated free-running SPD based on an InGaAs/InP NFAD for 1.5 $\mu$m Mie lidar system. We have optimized the key parameters of InGaAs/InP SPD and investigated the ringing effect of backscattering signal for lidar applications. After afterpulse and count rate corrections, the field lidar performance in terms of normalized $Pr^2$ versus range using the free-running InGaAs/InP SPD is considerably comparable to that of using a commercial SNSPD, with only $\sim$2\% relative error. Our technique provides a practical solution for single-photon detection dedicated to accurate lidar applications.

\section*{Funding}
This work has been financially supported by the National Basic Research Program of China Grant No.~2013CB336800, the National Natural Science Foundation of China Grant No.~11674307, and the Chinese Academy of Sciences.

\section*{Acknowledgments}
We would like to thank Wen-Hao Jiang, Chong Wang and Jiawei Qiu for useful discussions.

\end{document}